\documentclass[a4paper, 11pt]{easychair}

\usepackage{doc}
\usepackage{makeidx}
\usepackage{graphicx}
\usepackage{amsmath}
\usepackage{amsfonts}
\begin{document}

%
\title{ On the notion of ``von Neumann vicious circle'' coined by John Backus \\ {\small (February 2, 2016)}}


\titlerunning{  On the notion of ``intellectual von Neumann bottleneck'' }

%
\author{ Stanis\l aw Ambroszkiewicz 
}

\institute{
Institute of Computer Science, Polish Academy of Sciences \\ 
al. Jana Kazimierza 5, PL-01-248 Warsaw, POLAND,   \\
  \email{sambrosz@ipipan.waw.pl}
 }

\authorrunning{S. Ambroszkiewicz}

\clearpage

\maketitle


%
%

\pagestyle{empty}

\begin{abstract}
``The von Neumann vicious circle'' means that 
non-von Neumann computer architectures cannot be developed because of the lack of widely available and effective non-von Neumann languages. New languages cannot be created because of lack of conceptual foundations for non-von Neumann architectures.
The reason is that programming languages are high-level abstract isomorphic copies of von Neumann computer architectures. 
This constitutes the current paradigm in Computer Science. 
The paradigm is equivalent to the predominant view that computations on higher order objects (functionals) can be done only symbolically, i.e. by term rewriting. The paper is a short introduction to the papers arxiv.org/abs/1501.03043 and arxiv.org/abs/1510.02787 trying to break the paradigm by introducing a framework that may be seen as a higher order HDL (Hardware Description Language). 
 \end{abstract}


\section{Introduction} 

John Backus 1977 ACM Turing Award lecture ``Can Programming Be Liberated from the von Neumann Style'' \cite{Backus} is still a challenge.

``Von Neumann bottleneck'' (according to John Backus) is the limited throughput (data transfer rate) between the CPU and memory in the von Neumann computer architecture. It is much smaller than the rate at which the CPU can work. Due to the rapid hardware progress, CPU speed and memory size have been increasing much faster than the throughput between them. 

``The intellectual von Neumann bottleneck'' (according to Backus) is that programming can not go beyond   planning and detailing the enormous traffic of words through the von Neumann bottleneck.  

And finally, the ``von Neumann vicious circle''  is the fact that 
non-von Neumann computer architectures cannot be developed because of the lack of widely available and effective non-von Neumann languages. New languages cannot be created because of lack of conceptual foundations for non-von Neumann architectures.
The reason is that most of the  programming languages are high-level abstract isomorphic copies of von Neumann computer architectures. 

This constitutes the current paradigm in Computer Science. 
 To break the paradigm, John Backus proposed  function-level programming, where  
 high order mathematical structures (like functionals) are used as objects in programming. 
 Contrary to value-level programming where every elementary computation results in storing a  value (of primitive data type) in computer memory, function-level operates also on functionals as final results of computations. 
 
   Functional programming languages (like Haskell and F\#) are still von Neumann. 
Computation on the functionals is done there in a symbolic way by ``lazy evaluation''. That is, if a term (representing a functional) is to be fully evaluated to a value of a primitive data type, then  the term is rewritten to its normal form. Then, this normal form is compiled to the machine code of the von Neumann computer, and executed. 

It is worth to notice that the only non-von Neumann  programing languages are hardware description (and programing) languages (HDL)  like  VHDL and Verilog, where code can be compiled  onto FPGAs,  programmable integrated circuits. However, the languages are limited to the first order functions. In this context, the last publication of John von Neumann, {\em The Computer and the Brain}  \cite{vonNeumann} is of interest.

\section{ Functionals }  

The most intuitive methods of computing on the the natural numbers are recursive functions. The basic operation to define such  functions is the following schema of primitive recursion. 

Given two already defined functions $h$ ($k$-ary) and $g$ ($k+2$-ary), the new $f$ ($k+1$-ary) function is defined by the following equations. \\
$
f(1,x_1, … , x_k) = h(x_1, … , x_k) 
$\\
$f(n+1, x_1, … , x_k) = g(n, f(n, x_1, … , x_k), x_1, … , x_k)
$

Computation of $f$ for natural number $n$ consists in rewriting  the term $f(n)$ according to the above equations, where the right term can be rewritten as the left term, step by step starting from $n$, then, $n-1$, ... to 1. 

The following function may serve as a simple example.  $f: N \rightarrow N$, 
$f(1) =1$ and $f(n) = f(n-1)+n$  

Computation is done by the consecutive rewritings:  $f(5) = f(4) +5 =f(3) + 4 +5  =  f(2) +3 + 4 +5  = f(1) +2 +3 + 4 +5  = 1 + 2 + 3 + 4 +5$. 

The above primitive recursion schema may be abstracted to a higher level function (a functional of second order) that takes two  functions ($k$-ary and  $k+2$-ary) as its input, and returns, as its ouput, a  $k+1$-ary function. Let $R$ denote the functional. Then the above equations can be rewritten as \\
$
R(h,g)(1,x_1, … , x_k) = h(x_1, … , x_k)$ \\
$R(h,g)(n+1, x_1, … , x_k) = g(n, R(h,g)(n, x_1, … , x_k), x_1, … , x_k)
$

The term  $R(h,g)$ has no direct computational meaning unless it is fully evaluated, that is, for $R(h;g)(n, n_1, … , n_k) $, where $n, n_1, … , n_k $ are concrete natural numbers. 

 This kind of computations is known as term rewriting and is used ubiquitously in Computer Science. The above equations may be viewed as term rewriting rules. 
 


 To illustrate this symbolic way of computation (i.e. term rewriting), let us present some simple  examples from lambda calculus. 
 
Let $y$ and $x$ be variables, and $A$ and $B$ denote types,  such that 
$y:A\rightarrow B$ and $x:A$, so that the   term  $y(x)$ is of type $B$. 
 
Then, the term 
$\lambda y\lambda x .y(x)$ denotes a functional of type 
  $(A\rightarrow B) \rightarrow (A\rightarrow B)$ 


 Let the variable $z$ be of type  $B\rightarrow C$. 
 
 The term    
  $\lambda y\lambda z\lambda x.z(y(x))$ is of type $(A\rightarrow B) \rightarrow  (B\rightarrow C) \rightarrow  (A\rightarrow C)$  and denotes the functional  (say $F$) that for  $f: A\rightarrow B$  and  $g:B\rightarrow C$,  the term 
    $F(f)(g)$  is  of type $A\rightarrow C$),  such that $F(f)(g)(x) = g(f(x))$. 
    
The notation convention for application $()$ is that  $F(f)(g)(x)$ denotes the same as $((F(f))(g))(x)$. 
    
Intuitively, it is clear that the functional $F$ composes  any two functions of appropriate types. However, its computational content (meaning)  can be expressed only by term rewriting when it is fully evaluated to an object of type $C$. 
    
Note that application, i.e the operator ``$( )$'',  is  primitive in the terms. Can the composition be a primitive operation?

\subsection{ Higher order primitive recursion }  

Typed lambda calculus can be augmented by introducing combinators and corresponding requiting rules. One of them is Grzegorczyk Iterator \cite{Grzegorczyk1964}, that is a higher order version of the primitive recursion schema.   


It is (modulo currying)  introduced as  new symbol $R^A$ denoting functional of type 
 
$N \rightarrow (N\rightarrow (A\rightarrow A)) \rightarrow  ( (A\rightarrow A) )$ 

defined by the following equations. 
Let $a$, $c$ and $n$ be variables such that  $a:A$,  $c:N\rightarrow (A\rightarrow A)$, and  $n:N$, then 
 \\
$R^A(1)(c)(a) = a $\\
$R^A(n+1)(c)(a) = c(n)( R^A(n)(c)(a) )
$ 

Let the term
 $\lambda k\lambda c\lambda a .R^A(k)(c)(a)$   
be denoted by  $\bar{R}^A$ 
What is the functional denoted by this term? 
For 1 and any term $c$,  $\bar{R}^A(1)(c)$ denotes the identity function on $A$.  
For   $n >1$ and a term denoting a sequence $c$ of functions from $A$ into $A$,  the term  $\bar{R}^A(1)(c)$ denotes the composition of the first $n-1$ elements (functions)  of the sequence $c$. 
Again, if the composition were a primitive operation, then the computational meaning of the functional would be clear and simple. 
Since it is not the case, the computation must be done symbolically by term rewriting according to the above equations. 

%
%
 %
 %
   
 A lot of research was already done by prominent scientists 
  (Banach-Mazur, Rosa Peter,  G\"{o}del,  Grzegorczyk, Kleene, Kreisler, Scott, Platek, Girard, Reynolds, P. Martin-L\"{o}f),  to mention only some of them.

 The current paradigm in Computer Science is as follows. 
 Computations on higher order objects can be done only symbolically, that is, these objects can be represented and manipulated only using symbols.  Higher order objects (functionals) are identified with terms, whereas the computations on them are done by term rewriting. 
 
 
\section{ Higher order functionals as  hardware }

The approach proposed by C$\lambda$aSH http://www.clash-lang.org/ to realize higher order functionals is interesting. It goes 
 from the functional programming language Haskell and its high-level descriptions (syntax)  and via term rewriting (lazy evaluation as semantics) to a standard HDL.    Actually, after rewriting term (denoting a functional) fully to its normal form, that is, to imperative code, it is translated to a HDL. Hence, this approach is still unsatisfactory. For a recent survey of functional HDL, see \cite{gammie}. 

The notion of function as well as higher order objects (functionals) is based on the following more elementary notions. 
\begin{itemize}
\item type and object of type;
\item type constructors;
\item type of function and related input, function body, and output (the same for functionals;
\item  application of an object to the input of a function (functional), especially if 
 the input type is of higher order;
 \item composition of two functions (functionals). 
 \end{itemize}
Since the symbolic computation is challenged here, what is the alternative? 
Can these elementary notions listed above  be realized as hardware? 
Perhaps a solution is something like  dynamically configurable integrated circuits. 
However, FPGAs are limited to the first order functions, so that input as well as output of a FPGA circuit generally consist of  a fixed number of bytes.  




%
\begin{figure}[h]
	\centering
	\includegraphics[width=0.9\textwidth]{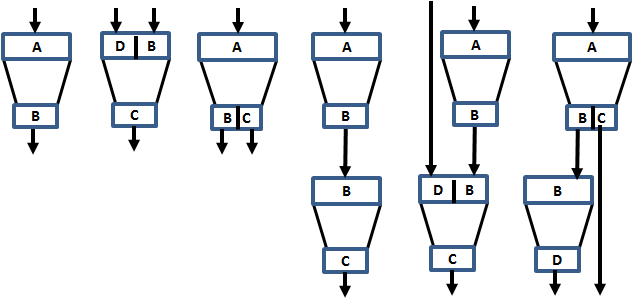}
	\caption{ Functionals and their compositions as establishing  links }
	\label{op}
\end{figure} 
\begin{figure}[h]
	\centering
	\includegraphics[width=0.7\textwidth]{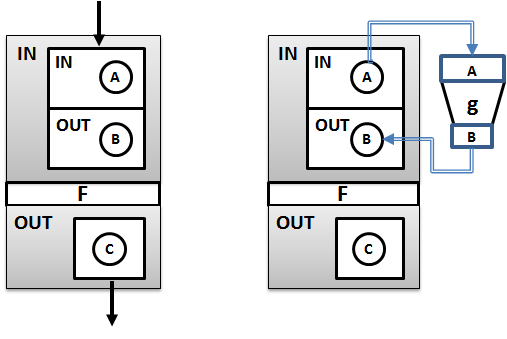}
	\caption{  Functional $F$ of type $(A \rightarrow B)\rightarrow C$. Input object  is a function $g$ of type $A\rightarrow B$. Application is done by establishing two links
 }
	\label{Ap}
\end{figure} 
\begin{figure}[h]
	\centering
	\includegraphics[width=0.99\textwidth]{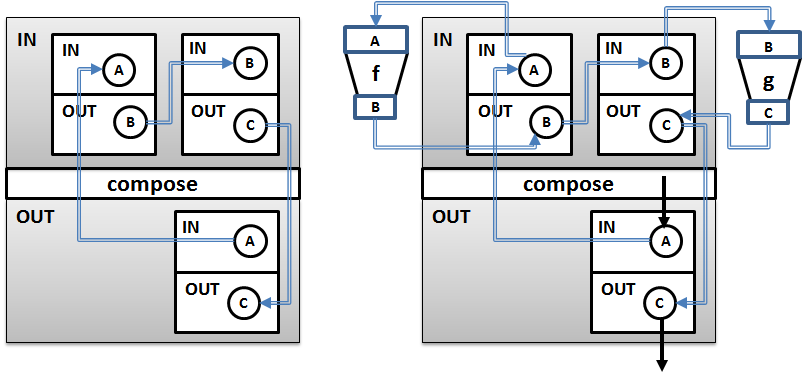}
	\caption{ The functional $Comp$ of type $((A \rightarrow B); (B \rightarrow C)) \rightarrow (A \rightarrow C)$. Input objects are $g$ of type $A\rightarrow B$ and $g$ of type $B\rightarrow C$.  When applied to $Comp$ the output object is of type  $A\rightarrow C$ }
	\label{Comp}
\end{figure} 

The following examples are taken from  \cite{TO}. 
Types are considered as boards of plugs and sockets. A function type consists of input type as a socket, function body, and output type as a plug, see Fig. \ref{op}. 

Composition of two function (functionals) is extremely simple and is a  link (connector)  between the output   type of the first function and the input type of the second function, see Fig. \ref{op}. The input type and output type must be the same, however, they play different roles, i.e. output type is a plug, whereas input type is a socket. 

Application of a higher order (where the argument is a functional) is also a configuration of appropriate links between boards, see Fig. \ref{Ap}. 

Composition as  a functional is a plug and socket board with the appropriate links, see Fig. \ref{Comp}. 


Generally, object of a higher type (functional) is a circuit consisting of  plugs and sockets and (dynamic) configuration of links between them. 
Computation on such objects is done by (dynamic) reconfiguration of the links in the objects.  
Note that the Grzegorczyk Iterator, as well as many other higher order functionals, can be realized as such circuits,  see \cite{TO}. 

The approach introduced in \cite{TO} and \cite{C} may be seen as a proposal of a high order hardware description language, perhaps extending standard HDL to higher order functionals realized as dynamically configurable integrated circuits.

\nocite{*}
 \bibliography{HOF}
\bibliographystyle{fundam}

\end{document}